# Edge-carboxylated Carbon Nanoflakes from Nitric Acid Oxidised Arc-discharge Material

Christoph G. Salzmann,*[a] Valeria Nicolosi[b] and Malcolm L. H. Green[a]

5 Carbon Nanoflakes (CNFs) with average diameters of ~30 nm have been prepared and isolated in bulk quantities by a single-step oxidation procedure using single-wall carbon nanotube arc-discharge material and nitric acid. The CNFs are predominately single, graphenic sheets containing a small number of internal defects. The edges are decorated with primarily carboxylic acid groups which allow facile chemical functionalisation and cross-linking of the fragments using multivalent
10 cations.

## Introduction

Two dimensional carbon materials have attracted considerable interest since the discovery that a single sheet of $sp^2$ carbon, graphene, can be isolated.[1,2] Graphene displays a variety of
15 remarkable physical properties such as high mobility of charge carriers,[3] mechanical stability[4] and thermal conductivity.[5]

The original preparation procedure for graphene relied on micromechanical exfoliation of graphite which yielded
20 relatively small amounts of sample.[2] Graphene layers have also been grown on metal substrates by using chemical vapour deposition techniques.[6] Very recently, graphene nanoribbons have been prepared by chemically 'unzipping' carbon nanotubes or by plasma etching carbon nanotubes.[7,8] Most
25 chemical procedures for the bulk production of graphene rely on the oxidation of graphite to graphite oxide in a first step, which requires very aggressive oxidation conditions, and subsequent chemical reduction by using, for example, hydrazine.[9-11]

30 For studies into the physical properties of graphene it is often advantageous to have as large sheets as possible. Regarding the chemical functionalisation, however, it may be more useful to have smaller fragment sizes which should disperse more readily in the reaction media. Furthermore, the
35 increased relative number of functional groups at the edges of smaller fragments should facilitate the structural characterisation of the product materials.

Here we show that predominately single-sheet Carbon Nanoflakes (CNFs) with average lateral diameters of about 30
40 nm can be produced in bulk quantities in a single step by oxidation of single-wall carbon nanotube arc-discharge material with nitric acid. The presence and location of functional groups has been investigated as well as the chemical reactivity and dispersibility in water and organic
45 solvents.

## Experimental

**Preparation and purification of materials.** Single-wall carbon nanotube (SWCNT) arc-discharge material (Carbolex AP grade or in-house produced material) containing a Ni/Y
50 catalyst[12] was subjected to a microwave treatment in 16% hydrochloric acid[13] followed by refluxing for 24 hours. In separate experiments, 500 mg of graphite (Aldrich, 282863), carbon nanopowder (Aldrich, 633100), catalyst free arc-discharge multi-wall carbon nanotube material (Aldrich,
55 412988) and the HCl treated SWCNT arc-discharge material were refluxed in 500 mL 9 M nitric acid for 24 hours. The filtrate of the SWCNT arc-discharge material was then neutralised under external ice cooling and vigorous stirring by slowly adding sodium hydroxide pellets. The formation of a
60 dark brown precipitate was observed which was isolated by filtration and stirred in 5 mL of a 0.2 M potassium cyanide solution for 24 hours. After filtration, the material was washed with KCN solution on the filter membrane, again stirred in KCN solution and recovered by filtration. The material was
65 then combined with 50 mL of deionised water and carefully neutralised by adding drops of 1 M hydrochloric acid under stirring. After filtration, the material on the filter membrane was washed with an excess of water and dried in a desiccator under vacuum (dry mass = 80 mg). For further purification,
70 some of the material was dispersed in 1 M formic acid, dialysed against 1 M formic acid solution and then against Millipore deionised water (10 mL Spectra/Por Float-A-Lyzer MWCO 2000). A dark brown powder was obtained after lyophilisation of the dialysed dispersion.

75 **Amidation reaction.** 5 mg of the purified material were dispersed in a mixture of 20 mL tetrahydrofurane (THF) and 3 mL oxalyl chloride, and heated at 40 °C under dinitrogen atmosphere for 18 hours. THF and oxalyl chloride were then evaporated under vacuum, and the residue was combined with
80 3 mL dry ethylenediamine and 20 mL dry THF. Upon stirring at 40 °C for four hours under dinitrogen precipitation was observed, and ethylenediamine and THF were removed under vacuum at 50 °C. The residue was stirred in water overnight, washed with an excess of water on a filter membrane and then
85 dried in a desiccator.

**Sample characterisations.** Optical absorbance spectra were recorded in quartz cuvettes (10 mm light path length) on a GBC spectrometer (Cintra 10). X-ray photoelectron spectroscopy was performed in an ion pumped UHV chamber
90 equipped with a VG nine channel CLAM4 electron energy analyzer using 300 Watt unmonochromated Mg Kα X-ray radiation. The analyzer was operated at constant pass energy of 100 eV for wide scans and 20 eV for detailed scans. A Digital Instruments Multimode Nanoscope Scanning Probe
95 Microscope was used for scanning tunnelling microscopy

(STM) and atomic force microscopy (AFM) in tapping mode. For STM and AFM analysis, the sample dispersions were dropped onto freshly cleaved 'highly oriented pyrolytic graphite' (HOPG) using a Laurell Technologies WS-400 spin-
5 coater (8000 rpm). The cleanliness of the substrate was ensured beforehand by using STM or AFM. For HRTEM observation, a dispersion of the sample was dropped onto a lacey carbon support grid (500 mesh). High-resolution transmission electron microscopy (HRTEM) images were
10 taken in a JEOL JEM-3000F FEGTEM (300 keV, point resolution 0.16 nm). Raman spectra were recorded on a Jobin Yvon Labram spectrometer using a HeNe laser (632.8 nm, 2 mW). ATR FT-IR spectra were obtained using a Nicolet 6700 spectrometer equipped with a goldengate ATR cell.

## Results and Discussion

A variety of carbon materials were subjected to an oxidative treatment by refluxing in 9 M nitric acid. In case of graphite, carbon nanopowder and the arc-discharge multi-wall carbon nanotube material colourless filtrates were obtained. The
20 reaction of the single-wall carbon nanotube (SWCNT) arc-discharge material with nitric acid on the other hand produced a dark orange filtrate (*cf.* Fig. 1).

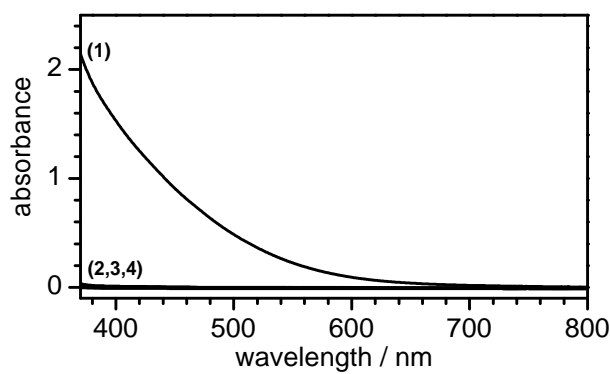

**Fig. 1** Optical absorbance spectra of the filtrates obtained after the
25 treatment of (1) arc-discharge single-wall carbon nanotube material, (2) graphite, (3) carbon nanopowder and (4) arc-discharge multi-wall carbon nanotube material in 9 M nitric acid for 24 hours at 100 °C. The curves of filtrates (2-4) overlap. Nitric acid absorbs strongly below ~370 nm and this spectral range is not shown.

30 After careful neutralisation of the filtrate with sodium hydroxide pellets a voluminous, dark brown precipitate was obtained which was isolated by filtration. Elemental analysis using X-ray photoelectron spectroscopy (XPS) showed that the sample contained carbon as well as oxygen, nickel(II) and
35 yttrium(III) (*cf.* Fig. 2(a), spectrum (1)). The nickel impurity was removed quantitatively by treatment with potassium cyanide solution (*cf.* spectrum (2)). Further purification using dialysis led to almost complete removal of yttrium from the sample (*cf.* spectrum (3)). After lyophilisation of the dialysed
40 dispersion, the sample was dispersible in water across the entire pH range as well as in other polar solvents such as tetrahydrofurane, dimethylformamide and methanol.

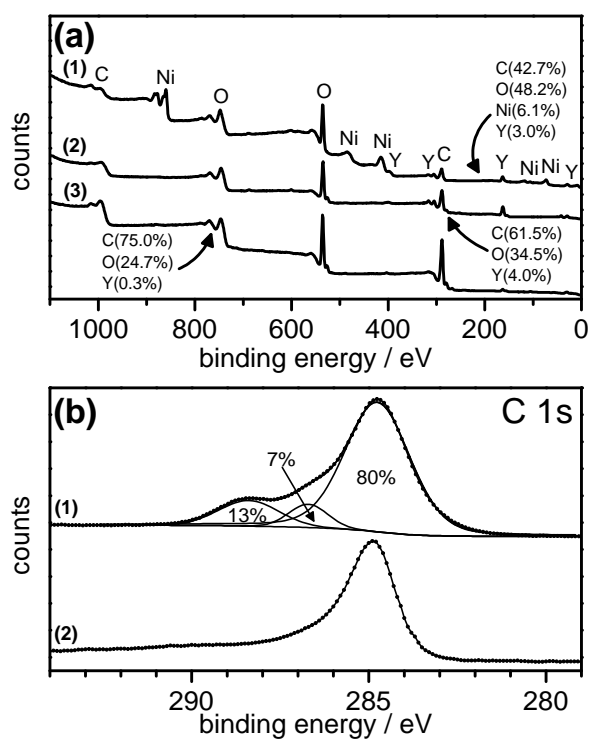

**Fig. 2** (a) X-ray photoelectron spectroscopy survey scans of the material
45 (1) after precipitation, (2) after treatment with KCN solution and (3) after dialysis. The elemental composition of the samples is given in atom percent for each spectrum. (b) High-resolution spectra of the C 1s region of the purified sample (1) before and (2) after annealing at 850 °C under argon. Fitted peaks and relative area percentages are shown for spectrum
50 (1).

The high-resolution XP spectrum of the C 1s region of the purified sample showed that a large portion of the carbon atoms were unoxidised after the nitric acid treatment as indicated by the binding energy of 284.7 eV of the main peak
55 (*cf.* spectrum (1) in Fig. 2(b)). This illustrates the relatively 'weak' oxidation power of 9 M nitric compared to, for example, the conditions necessary for the formation of graphite oxide from graphite (*e.g.* $NaClO_3$/fuming $HNO_3$) after which the peak around 284.7 eV was either greatly
60 diminished or not present anymore (*cf.* Fig. 8 in ref. 14). The weak peak at 286.7 eV is attributed to C(I) or C(II) species, and the peak at 288.4 eV is commonly assigned to C(III) species such as carboxylic acids, anhydrides or lactones.[15, 16] After heating the sample at 850 °C under argon the peaks at
65 higher binding energies disappeared, and the spectrum typical for graphitic carbon with a peak centred at 284.9 eV and tailing on the high-energy side was obtained (*cf.* spectrum (2)).[15, 17] Peak asymmetry, which is almost impossible to fit, may also present for the main peak in spectrum (1), and the
70 peak areas of the C(I) or C(II) species in particular but also of the C(III) carbon species may therefore be lower than the values obtained from peak fitting using symmetric peaks. Consequently, the peak area of graphitic carbon is expected to be slightly higher than 80 %.

75 The FT-IR spectrum of the purified sample (*cf.* spectrum (1) in Fig. 3(a)) shows a broad peak centred at 1714 cm$^{-1}$ which is about the ν(C=O) stretching frequency expected for carboxylic acids groups (1700 – 1725 cm$^{-1}$) or lactones (~1735 cm$^{-1}$).[18] Anhydrides typically show higher stretching frequencies. For example, 1779 cm$^{-1}$ is observed for 1,4,5,8-naphthalenetetracarboxylic dianhydride.[19] Anhydride groups are therefore probably not present in the sample, at least not in significant quantities. The lower frequency peak at 1600 cm$^{-1}$ is assigned to aromatic ν(C=C) stretching modes. After annealing the sample at 850 °C under argon the ν(C=O) peak disappears completely as a consequence of decarboxylation (*cf.* spectrum (2)). The purified sample was also reacted with oxalyl chloride and then with ethylenediamine (*cf.* Experimental). Carboxylic acid groups (COOH) are expected to form amides under these reaction conditions, and the spectrum of the sample after the reactions showed that the ν(C=O) mode had shifted to 1652 cm$^{-1}$ which is in accordance with the formation of an amide (*cf.* spectrum (3)).[18] XPS analysis also confirmed the presence of nitrogen in the sample after the reactions (not shown). From this it is concluded that a substantial fraction of C(III) species present in the sample are carboxylic acid groups.

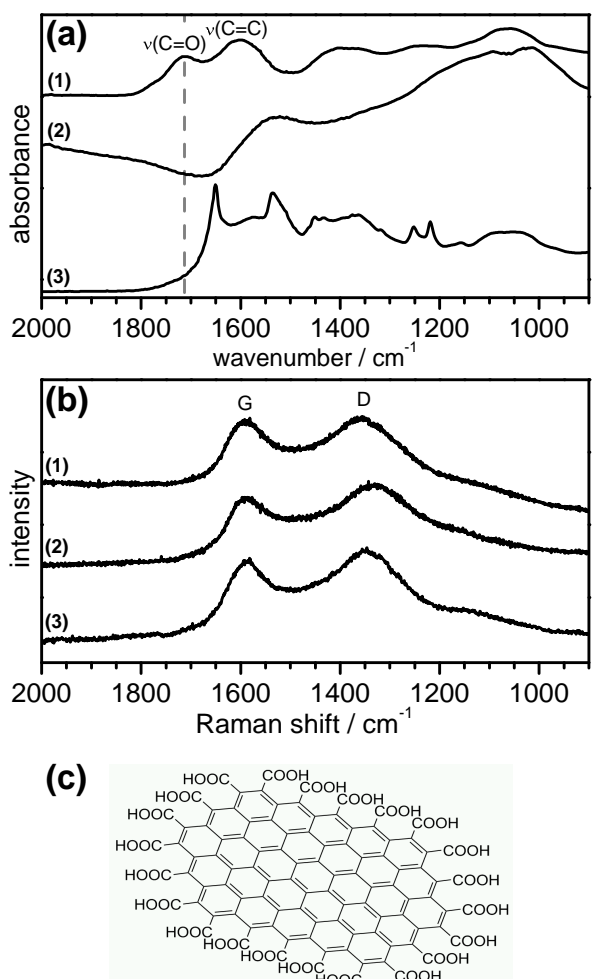

**Fig. 3** (a) FT-IR spectra of (1) the purified sample, (2) after annealing at 850 °C under argon and (3) after the amidation reaction. (b) Raman spectra of the same samples. (c) Idealised chemical structure of a small edge-carboxylated Carbon Nanoflake.

Raman spectroscopy showed the presence of sp$^2$ carbon in all samples as indicated by the G-band ('graphitic band') at ~1590 cm$^{-1}$ and the D-band ('disorder band') at ~1360 cm$^{-1}$ (*cf.* Fig. 3(b)). The high-temperature annealing under argon shifted the D-band to ~1330 cm$^{-1}$. The D-band in carbon materials is generally associated with the presence of 'disorder' such as defects or simply nanoscale dimensions.

The spectroscopic data leads to the conclusion that the sample consists of carboxylated Carbon Nanoflakes (CNFs) similar to the one shown schematically in Fig. 3(c). Scanning tunnelling microscopy (STM) (*cf.* Fig. 4(a)) and transmission electron microscopy (TEM) (*cf.* Fig. 5(a)) showed, however, that the fragments were much larger than the one depicted in Fig. 3(c). The average lateral length, measured along the longest distance across the fragments, is 30 ± 9 nm as determined from the STM data. STM step-height analysis showed that the largest fraction of the fragments (~70%) are about 0.7 nm in height (*cf.* Fig. 4(b,c)). Only a few fragments had about twice or three times this step-height suggesting that the majority of the fragments consist of single layer sheets and that only a few double and triple layered CNFs are present in the sample.

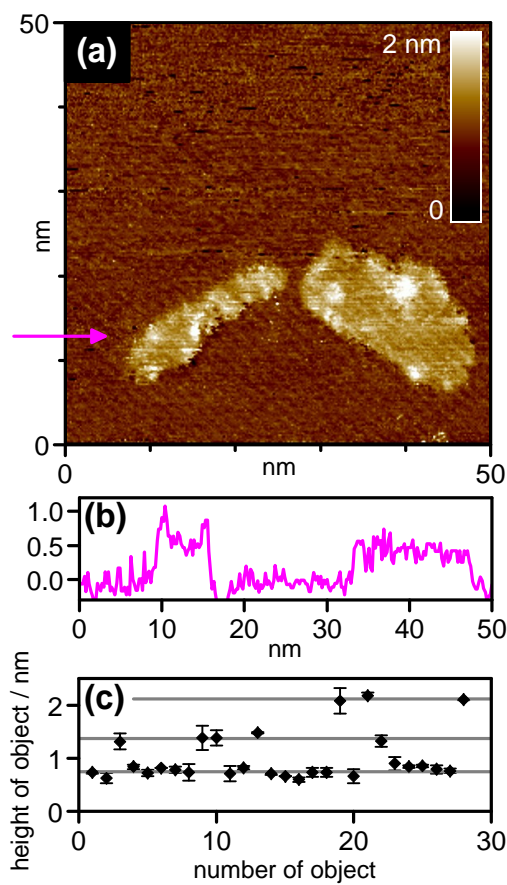

**Fig. 4** (a) Scanning tunnelling microscopy image of two Carbon Nanoflakes. (b) Height profile along the direction indicated in (a). (c) Height analysis of the CNFs. Horizontal lines indicate the step-heights for single, double and triple layer fragments.

The existence of single sheets was also shown by using normal-incidence selected area electron diffraction measurements which showed the pattern expected for a single sheet of graphene with the innermost reflections being the most intensive ones (*cf.* Fig. 5(c)).[20, 21] High-resolution TEM of the CNFs showed the regular hexagonal structure expected for graphene. Only a small number of internal defect 'islands', such as the one indicated by the arrow in Fig. 5(b), could be found within the graphene sheets.

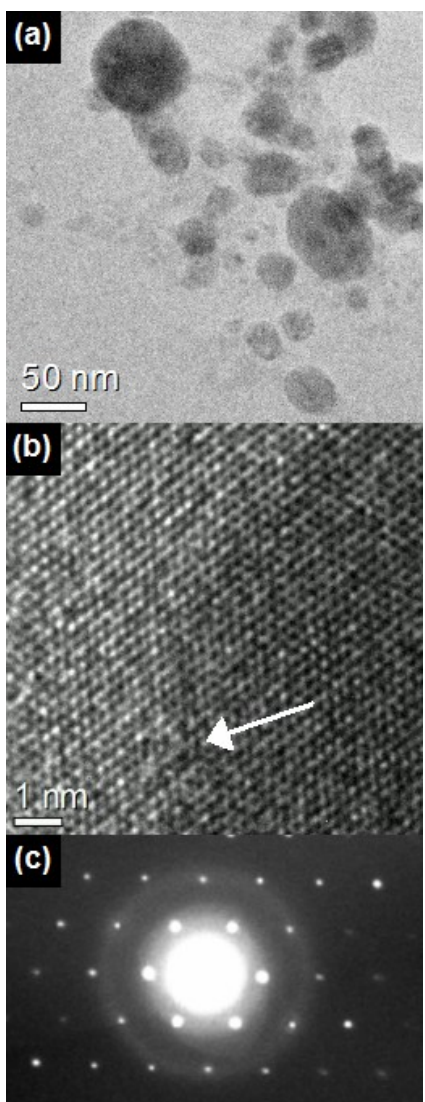

**Fig. 5** (a) Transmission electron microscopy image of Carbon Nanoflakes deposited onto a lacey carbon support grid. (b) High-resolution image of the interior of a Carbon Nanoflake. The arrow indicates a structural defect. (c) Electron diffraction pattern of the interior of a CNF.

The binding of COOH groups onto pristine and defective graphene was investigated computationally.[22, 23] The defects observed here are, however, larger than the ones investigated computationally and contain therefore most likely several functional groups – assuming that the defects do not form due to the exposure of the sample to the electron beam. In general, the mechanism for the formation of internally carboxylated graphene would have to involve the transformation of a carbon atom already existing within the graphene sheet to a COOH group. The adsorption of COOH onto pristine or defective graphene seems chemically unlikely.

The small number of defects within the sheets suggests that the majority of the COOH groups are decorating the edges of the sheets as shown schematically in Fig. 3(c). In order to investigate this further, 25 μL of a 1 mg mL$^{-1}$ CaCl$_2$ solution were added to 1 mL of a 1 mg mL$^{-1}$ CNF dispersion resulting in a final Ca$^{2+}$ concentration of ~0.2 mmol L$^{-1}$. The divalent Ca$^{2+}$ cations are expected to be complexated by the carboxylic acid groups at neutral pH,[24] thereby cross-linking CNFs together. Indeed, complete precipitation was observed within about 30 minutes. Similar cross-linking using Mg$^{2+}$ or Ca$^{2+}$ cations has in fact been shown to improve the mechanical properties of graphene oxide papers.[25]

After precipitation, the sample vial was subjected to ultrasonication using a bath sonicator for about 3 minutes which led to complete redispersion of the sample. A drop of the dispersion was then quickly dropped onto a freshly cleaved, rotating HOPG substrate in a spin coater. Analysing the surface using atomic force microscopy (AFM) revealed that the CNFs were aggregated 'edge-to-edge' and formed agglomerates of two, three or four fragments under these conditions. Fig. 6 shows such an agglomerate which consists of three cross-linked CNFs. Apart from CNF-agglomerates, it was also found that the CNFs decorate the step defects of HOPG suggesting that the steps may also be terminated by some functional groups, such as COOH groups, which can bind Ca$^{2+}$ ions. The fact that the fragments join together 'edge-to-edge' and do not 'pile up' strongly suggests that the majority of the COOH groups are decorating the edges of the CNFs.

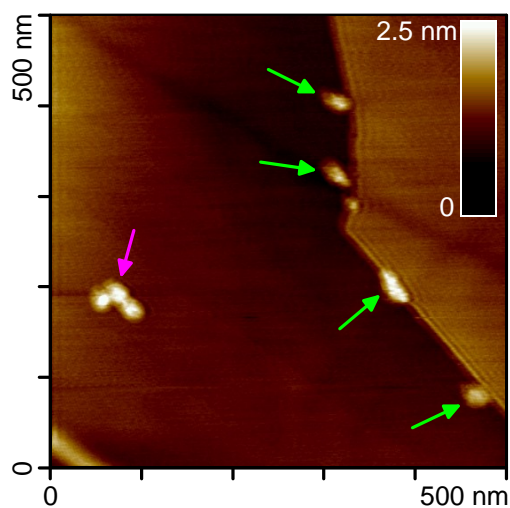

**Fig. 6** Atomic force microscopy image of a freshly ultrasonicated Carbon Nanoflake dispersion containing Ca$^{2+}$ deposited onto HOPG. Arrows indicate the aggregation of Carbon Nanoflakes with each other or at a step defect.

CNF dispersions were also found to precipitate upon addition of YCl$_3$ solution, and this is most likely the reason why precipitation is observed upon neutralising the filtrate of the initial nitric acid solution in which some yttrium(III) is present from the Ni/Y catalyst (*cf.* Experimental and Fig. 2(a)). This illustrates how the presence of small amounts of multivalent cations can compromise the dispersion properties of carbon materials containing carboxylic acid groups.

Subjecting the SWCNT material to a second treatment with nitric acid led again to the formation of a dark orange filtrate. However, in order to precipitate the material upon neutralisation, it was necessary to add some $CaCl_2$ or $YCl_3$ solution before neutralisation.

Arc-discharge SWCNT materials are well known to contain considerable fractions of 'amorphous carbon'.[26, 27] The fact that the CNFs can be isolated after the oxidative treatment of the arc-discharge material using nitric acid suggests that graphitic fragments of similar dimensions may already be present in the as-made material. The fragments are perhaps interconnected by carbon species which can be oxidised by nitric acid which then liberates the CNFs. It has been shown previously that as-made SWCNTs are quite resilient towards oxidation with 9 M nitric acid.[28] The dimensions of the CNFs make it seem unlikely that the fragments originate from the decomposition reaction of SWCNTs. It has been shown previously that some of the CNFs adsorb onto SWCNT materials, and they were identified as the main 'carriers' of the COOH functionality after treatment of as-made SWCNT materials with nitric acid.[28-30] Very recent work has shown that the arc-discharge process can be optimised towards the formation of 2-4 layer graphene in the presence of hydrogen.[31]

The oxidative treatment of the multi-wall carbon nanotube arc-discharge material with nitric acid did not lead to the formation of CNFs as it can be seen from the optical absorbance spectrum of the filtrate shown in Fig. 1. The multi-wall material does not contain a Ni/Y catalyst which shows that the presence of the metal catalyst is not only important for forming SWCNTs but also the graphitic fragments. It is seems possible that the graphitic fragments form as a side-reaction to the SWCNT formation. Unlike single-wall carbon nanotubes, which can grow up to lengths of several micrometers, the growth process of the fragments seems to stop much earlier as reflected by the smaller dimensions of the CNFs.

## Conclusions

Flakes of graphitic carbon with average lateral diameters of ~30 nm have been prepared in bulk quantities. The Carbon Nanoflakes (CNFs) consist predominantly of single-sheets and show a small number of defects within the sheets. Unlike the preparation of chemically modified graphene from graphite which, in most cases, requires the oxidation of graphite to graphite oxide first, the preparation of CNFs takes place in a single oxidation step using arc-discharge single-wall carbon nanotube material and nitric acid. The edges of the graphene fragments are decorated with mainly carboxylic acid groups (COOH). The CNFs show excellent dispersion properties in water and other polar solvent in the absence of multivalent cations, and the COOH groups can be easily functionalised using carboxylate chemistry. The dispersion and chemical reactivity properties of this material make them ideally suited for further studies into the chemical functionalisation of oxidised graphene edges. Applications in which the CNFs act as a 'junction box' mediating electron transfer between redox-active groups linked to the edges of the CNFs are anticipated.

## Acknowledgements

We thank Dr R Jacobs for help with scanning tunnelling microscopy, Dr A Crossley for help with X-ray photoelectron spectroscopy, and Drs J M Brown, K A Vincent and J. A. Raskatov for discussions. We are grateful to the Austrian Academy of Science for an APART grant (C.G.S.) and the EPSRC for granting access to the NCESS facility (EP/E025722/1).

## Notes and references

[a] *Department of Chemistry, Inorganic Chemistry Laboratory, University of Oxford, South Parks Road, OX1 3QR Oxford, United Kingdom. Fax: 0044 (0) 1865 2 72690; Tel: 0044 (0) 1865 2 72645; E-mail: christoph.salzmann@chem.ox.ac.uk*
[b] *Department of Materials, University of Oxford, Parks Road, OX1 3PH Oxford, United Kingdom*